# Structures and energies of computed silicon (001) small angle mixed grain boundaries as a function of three macroscopic characters

**Author information**

Wei Wan[1, 2] and Changxin Tang[*, 2]

[1] *Institute for Advanced Study, Nanchang University, Nanchang, China*

[2] *Institute of Photovoltaics, Nanchang University, Nanchang, China*

[*] *Corresponding author, Email address:* tcx@ncu.edu.cn

## Abstract

Understanding how dislocation structures vary with grain boundary (GB) characters enables accurate controls of interfacial nano-patterns. In this atomistic study, we report the structure-property correlations of Si (001) small angle mixed grain boundaries (SAMGBs) under three macroscopic GB characters (tilt character, twist character, and an implicit rotation character between them). Firstly, the SAMGB energies are computed as a function of tilt angle, twist angle and rotation angle, based on which a revised Read-Shockley relationship capable of precisely describing the energy variations span the three-dimensional GB character space is fitted. Secondly, GB structural transitions from dislocation to amorphous structures are given as a function of tilt angle, twist angle and dislocation core radii. The proportion, topology and structural signatures of different SAMGB types defined from the ratio between the tilt and twist angles are also presented. Thirdly, by extracting the transformation of metastable SAMGB phases, the formation mechanisms of SAMGB structures are characterized as energetically favorable dislocation glide and reaction, from which the dislocation density function is derived. The relevant results about SAMGB energies and structures are validated and supported by theoretical calculations and experimental observations, respectively.

## 1. Introductions

Grain boundaries (GBs) are interfaces separating individual crystals with unique impacts on multiple material properties, such as mechanical strength [1, 2], impurity segregation [3], corrosion susceptibility [4] and electronic properties [5]. Since silicon became one of the most frequently used materials in the emerging semiconductor and photovoltaic industries, there raised demands of grain boundary engineering (GBE) to improve the material properties by controlling the population of desired GB types [6–8]. Unfortunately, much of our knowledge about GBs and incremental attempts on material properties were focused on FCC metals with low stacking fault energies [9, 10]. To enable the high performance of related silicon materials like multi-crystalline silicon and nano-crystalline silicon [11], more detailed information on the structure-property correlations of different GB types is required.

The structures of a given GB are jointly defined by five well-known macroscopic degrees of freedom (DOFs), and many microscopic DOFs of the interfacial atoms. The macroscopic DOFs are also known as GB characters, three of them characterize the misorientation between two crystals and the other two describe the orientation of the GB plane. The microscopic DOFs could be seen as varying the GB atoms to form a multiplicity of GB phases or metastable structures [12–14]. Various metrics were proposed to characterize the atomic structures of GBs that span macroscopic and microscopic DOFs, such as the structural unit model [15, 16], polyhedral unit model [17] and local atomic bond environments for machine-learning representations [18–22]. There are also many important frameworks over various GB properties. The most notable one should be the Read-Shockley relationship [23], which predicts GB structures and energies as a function of misorientation angle (e.g., tilt or twist angle) with strong reliability over the years. Other noteworthy frameworks include the broken-bond energy model [24] contributed by D. Wolf and a closed-form GB energy function of FCC metals [25] from Bulatov et al. The latest study about GB universality [26, 27] has demonstrated that the GB structures and energy trend are more sensitive to the geometry (i.e., macroscopic structural descriptions) than different material types, and thus these mentioned metrics and structural-property correlations have certain transferability. As the macroscopic DOFs form a vast and complex five-dimensional GB character space, they were often simplified down to one DOF (misorientation angle) based on the GB character with simple geometry like tilt and twist, by completely ignoring the orientation of the GB plane and misorientation axis [9, 28–32]. Although the simplification is quite useful to deal with the structure-property correlations, it still represents an incomplete characterization of possible GB characters.

It should be noted that the GB character is not limited to the tilt or twist from the simplification. An arbitrary GB structure is more likely to be the mixed tilt-twist character with tilt and twist components simultaneously. In that case, the independent tilt and twist characters could be treated as two decomposed components of the mixed tilt-twist character, which forms a two-dimensional GB character space that complicated classical Coincidence Site Lattice (CSL) descriptions. Nearly two decades ago, Morawiec and Glowinski [33] had already proposed a theoretical characterization of the mixed GB character and highlighted that such character should be investigated by its decomposition into tilt and twist components. But until now, our cognition of the mixed GB character is still confined to the theory as there are few reports on their structure-property correlations. For years, the simplest one-DOF tilt and twist GBs and the orientation of the GB plane were the major focus, both experimentally and numerically [29–32, 34–38]. While high-resolution transmission electron microscopy (HRTEM), electronic beam induced current (EBIC) and other experimental methods had gained great success for small angle grain boundaries (SAGBs) and high angle tilt GBs [34–40], atomistic simulations were still believed particularly insightful for modelling relatively complex GB structures (e.g., twist) that are difficult for experimental assessments [41–43]. From the pioneering work of Olmsted et al. [44, 45] and the fundamental zone of the orientation of the GB plane [46, 47] to recent statistical examination [48] of the entire five-dimensional GB character space, most atomistic approaches have ignored or simplified this complex GB character. Undeniably, as a two-dimensional subset of the five-dimensional GB character space, the mixed GB character still requires a huge amount of work to fully understand. Nevertheless, what we wish to highlight here is a small but interesting portion that frequently appeared in the bonded thin films [49, 50]: the small angle mixed grain boundaries (SAMGBs) comprised of dislocation structures.

The SAMGBs are a part of the general SAGBs defined from the simplification, and they should contain the characteristics from their tilt and twist components, which are often modelled as infinite dislocation arrays and screw dislocation networks, respectively. For example, it has been experimentally and numerically shown that the small angle twist GB is a quadrate screw dislocation network [50–52] in Si (001) and then varies to hexagonal in Si (111) shuffle and triangular in Si (111) glide [53, 54]. The presence of zigzag dislocation segments [55] in Si (001) small angle twist GBs indicated that the GB observed there may not belong to pure twist. It also found that even the small angle symmetric tilt GBs in silicon have dependence and multiplicity on their dislocation structures and energies [56, 57]. More specifically, since the formation of an undesired screw dislocation network introduced by a small twist angle was almost inevitable in the direct wafer bonding [49], the SAMGBs were often observed as a complex dislocation network perturbed by a mixed dislocation array contributed by a small tilt angle [50]. The knowledge about dislocation structures of SAGBs from these reviewed studies has its uses. On the one hand, these complex dislocation networks are usually known as nano-patterns [51], which provide a template for self-assembled nanostructures and hence control the performance of thin films [58, 59]. On the other hand, the decrease of edge dislocation density in the crystalline silicon materials (e.g., multi-crystalline, mono-crystalline and mono-like silicon) through depopulating designated SAGB types could benefit the photoelectric conversion of solar cells as those SAGBs with edge dislocations were proven to be the most electronically active defect in silicon [60–62]. Therefore, advancing the understanding of the structures and properties of unknown SAMGBs is highly desirable. The Continuum-based theoretical methods may be a possible way to solve the problem, but there are limitations to their uses. For example, the Frank-Bilby Equation (FBE) that connects the macroscopic and microscopic GB DOFs is often used to deal with the interfacial dislocation arrays [63–65], but it is unable to uniquely determine those complex SAMGB structures. Beyond the FBE, other theoretical approaches based on the Continuum model [66] or analytical Peierls-Nabarro model [67] with the capacity to provide energetic properties were established for arbitrary SAMGB by approximately treating it as infinite dislocation arrays and ignoring possible dislocation glide and reaction. However, in most cases, a direct correlation (e.g., Read-Shockley relationship) connecting misorientation angles and properties without presuming SAMGB structures is desired, which requires further revealing the formation mechanisms of SAMGBs. Generally speaking, the SAMGB can be investigated by its decomposition, while its structure-property correlations and formation mechanisms are two raising questions waiting to be solved.

A key aspect of our approach is that all SAMGBs are studied as a function of their tilt and twist characters simultaneously without any simplification, which means examining the variations of structures and energies in the two-dimensional GB character space. We also try to characterize the formation mechanisms of their structures considering possible dislocation glide and reaction. This work is based on the atomistic simulation of the simplest (001) of three major planes of silicon, where many experimental works were carried out [49–51, 55, 68, 69]. Thus, the validity of our work could be directly verified by in-situ TEM observations and generalize the results to a wider class of diamond lattice materials (e.g., diamond) through the universality of GB structures.

## 2. Methodologies

### 2.1. Geometry and potential

An arbitrary mixed GB character could be geometrically expressed by using three variables $u$, $v$ and $w$, and three orthogonal unit vectors $i$, $j$ and $k$. Initially, the orientations of two perfect crystals are $i$, $j$ and $k$ along the X, Y and Z axes, respectively. Then their orientations vary to

$$\begin{aligned} i_{upper} &= ui + vj + wk \\ j_{upper} &= -vi + uj \\ k_{upper} &= -uwi - vwj + (u^2 + v^2)k \end{aligned} \quad (1)$$

for the upper (Z+ direction) crystal and

$$\begin{aligned} i_{lower} &= ui - vj - wk \\ j_{lower} &= vi + uj \\ k_{lower} &= uwi - vwj + (u^2 + v^2)k \end{aligned} \quad (2)$$

for the lower (Z– direction) crystal, after introducing a twist angle $\phi = 2\times\arctan(v/u)$ in the XY plane and a tilt angle $\theta = 2\times\arctan(w/u)$ in the XY plane with the tilt symmetry axis Y. Figure 1 summarizes the geometry of the mixed GB character. The total misorientation angle $A_t$ between the two crystals for the mixed GB character is determined by:

$$A_t = \arccos\left(\frac{u^2 - v^2 - w^2}{u^2 + v^2 + w^2}\right) \quad (3)$$

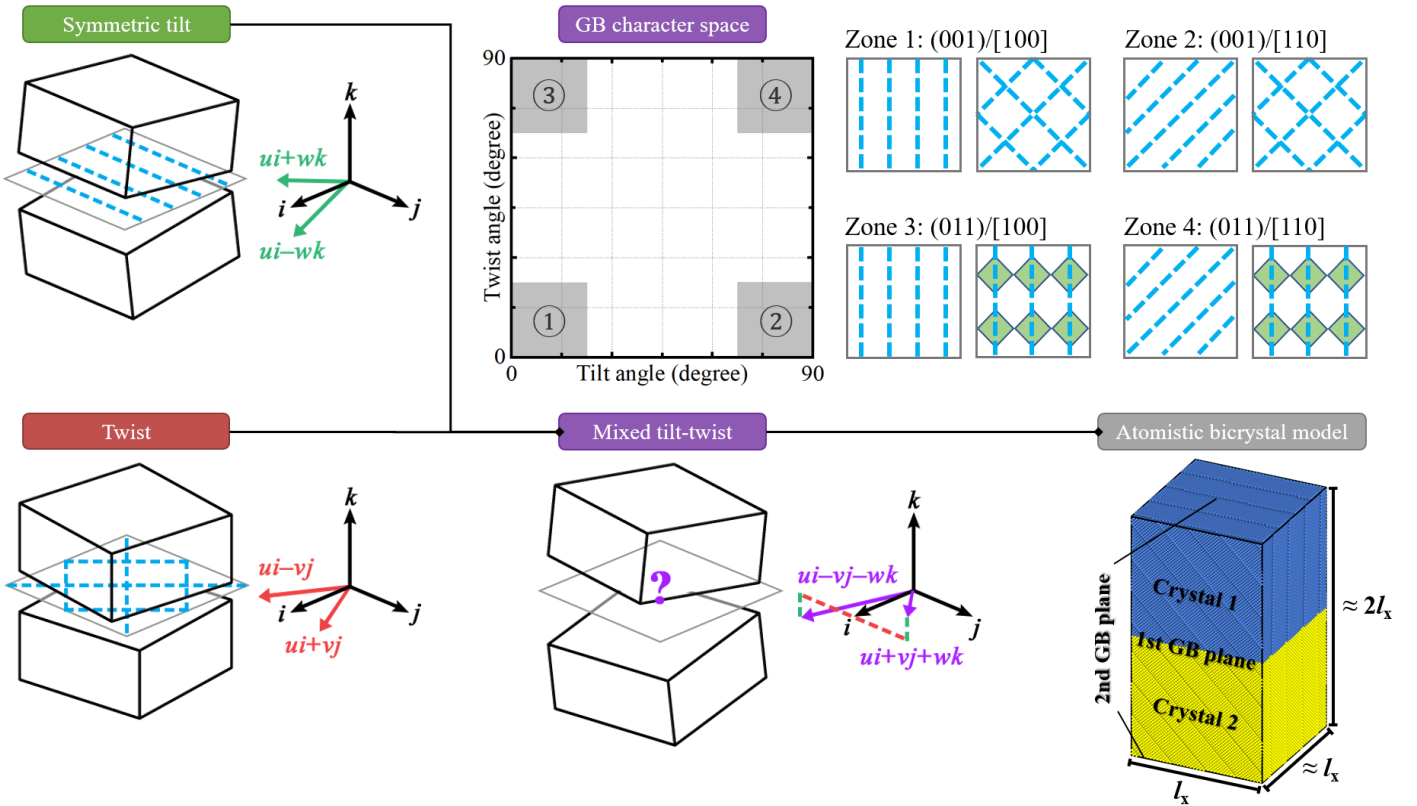

**Figure 1.** Schematic representations of geometry and modelling of the symmetric tilt, twist and mixed tilt-twist GB characters. In the two-dimensional GB character space, small angle mixed GBs that comprised dislocation structures could be found in four zones, which are marked with numbers 1-4. For each zone, we give its GB plane normal (e.g., (001)) and tilt symmetry axis within the GB plane (e.g.,[100]) to uniquely characterize a mixed GB character. The left one is the tilt component and the right one is the twist component. The blue dash line denotes dislocations and the green rhombus denotes stacking faults for (011) small angle twist GBs.

This work focuses on the small angle portions of mixed GB characters on the (001) plane in diamond silicon, where the orientations $i$ = [100], $j$ = [010] and $k$ = [001]. The one-DOF symmetric tilt and twist GBs are crystallographic equivalents for misorientation angles $A$ and 90°-$A$ due to the fourfold symmetry <100> axis. However, only the misorientation angles $A$ and 180°-$A$ are equivalent for the mixed GB due to its twofold symmetry, which results in four different small angle zones in Figure 1. Here we only consider examining Zone 1 with <100> normal that falls in the topic. In simplification, it is usually believed that one-DOF GBs are comprised of dislocation structures when $A$ is below 25° [70]. Therefore, the important small angle misorientations of the mixed GB character are expected in the $\theta$ range 0° < $\theta$ < 25° and $\phi$ range 0° < $\phi$ < 25°. Within this fundamental two-dimensional GB character space, 121 (001)/[100] mixed GB characters are sampled, which are formed by the free combinations between 11 (001)/[100] symmetric tilt GBs (range from the lowest tilt angle $\theta_{lo}$ = 1.15° to the highest tilt angle $\theta_{hi}$ = 22.62°) and 11 (001) twist GBs (range from the lowest twist angle $\phi_{lo}$ = 1.15° to the highest twist angle $\phi_{hi}$ = 22.62°). Table 1 shows the details of the sampled tilt and twist components.

**Table 1.** CSL representations of the used tilt and twist components.

| ID | Tilt components | | | Twist components | | |
|---|---|---|---|---|---|---|
| | Σ | Orientation | $\theta$ (Degree) | Σ | Orientation | $\phi$ (Degree) |
| 1 | 10001 | [100 1 0] | 1.15 | 10001 | [100 0 1] | 1.15 |
| 2 | 2501 | [50 1 0] | 2.29 | 2501 | [50 0 1] | 2.29 |
| 3 | 313 | [25 1 0] | 4.58 | 313 | [25 0 1] | 4.58 |
| 4 | 2509 | [50 3 0] | 6.87 | 2509 | [50 0 3] | 6.87 |
| 5 | 629 | [25 2 0] | 9.15 | 629 | [25 0 2] | 9.15 |
| 6 | 101 | [10 1 0] | 11.42 | 101 | [10 0 1] | 11.42 |
| 7 | 317 | [25 3 0] | 13.69 | 317 | [25 0 3] | 13.69 |
| 8 | 2549 | [50 7 0] | 15.94 | 2549 | [50 0 7] | 15.94 |
| 9 | 641 | [25 4 0] | 18.18 | 641 | [25 0 4] | 18.18 |
| 10 | 2581 | [50 9 0] | 20.41 | 2581 | [50 0 9] | 20.41 |
| 11 | 13 | [5 1 0] | 22.62 | 13 | [5 0 1] | 22.62 |

Various semi-empirical interatomic potentials are available for diamond silicon. We employ a newly modified potential from Kumagai et al. [71] This many-body potential follows the Tersoff formalism [72] and limits interactions to the first-neighbour shell, with precise reproductions of a series of density functional theory (DFT) results including elastic constants and stacking fault energies of silicon (Table A1 and Figure A1 in the Appendix). These features guarantee an accurate and reasonable representation of diamond silicon and its dislocations, at the same time, a low computational cost allowing detailed sampling of large-scale GB structures.

## 2.2. Atomistic Simulation Procedure

GB structures with minimum excess energy are generated in a bicrystal model by molecular mechanics simulations from our in-house LAMMPS [73] code "GB-Gen" [50], which is mainly based on the work of Tschopp et al. [74]. Their geometry is set according to Figure 1 and equations (1) and (2). Periodic conditions are used in the X, Y and Z axes. The box size along Z is approximately two times of box size along X to minimize the interaction between two generated GBs.

Then, the two perfect crystals are placed relative to each other to sample initial configurations of atom positions in the Cell of Non-Identical Displacement (CNID), which is defined by the corresponding Displacement Shift Complete (DSC) lattice vectors. 20 X positions and 20 Y positions are used, which means the sampling increments are fractions 0.05, 0.1 . . . 0.95 of the DSC vectors. 20 different placement positions of the GB plane are considered. However, atom removal after proximity is not applied for those with definite dislocation structures as dislocations are not sensitive to that. These sampling operations (varying with the misorientation angles) for 121 unique mixed GB characters create over 0.8 million initial configurations for the subsequent conjugate-gradient energy minimization. To the author's knowledge, this number is little than the computational dataset of Homer et al. [48] covering the five-dimensional GB character space, but the total computation cost is quite close due to the average simulation scale (~ 2 million atoms).

To eliminate unreasonable dislocation configurations, stable and nearest metastable 0K GB structures are quenched under 1000K and 1atm, which enable the rearrangement of interfacial atoms and further activate potential dislocation dynamics. The quenching process was proved necessary to find ground-state GB configurations at finite temperatures in earlier numerical efforts [9], and experimental processes similar to that have also been adopted in the direct bonding technics [49–51] while processing interfacial dislocation structures. After the quenching process, the GB structures are thermalized to 300K and 1atm using the Berendsen ensemble [75]. The potential energy, interatomic force and Virial stress tensor of each atom are collected to calculate the GB energies.

## 2.3. Grain boundary characterization

The GB energy $E_{GB}$ is computed from the relative cohesive energy difference between atomic potential energies $E_i$ and $E_{coh}$ (measure under 300K and 1atm) for the model with and without the GB, respectively as:

$$E_{GB} = \frac{\sum_{i}^{N}(E_i - E_{coh})}{2A_{GB}} \quad (4)$$

where $N$ is the atom count of the bicrystal system and $A_{GB}$ is the GB area. For SAMGBs, $E_{GB}$ can be divided into dislocation core energy $E_c$ and dislocation strain energy $E_s$, where $E_s$ is computed through atomic Virial stress $\sigma_i^{atom}$ following:

$$E_s = \sum_{i}^{N} \frac{1}{2} V_i^{atom} \left(\sigma_i^{atom}\right)^T S \sigma_i^{atom} \quad (5)$$

where $V_i^{atom}$ is the volume of atom $i$. Superscript $T$ represents tensor transpose. Elastic compliance tensor $S$ is obtained by the inverse of the elastic stiffness tensor. $E_c$ is unable to compute directly, and thus it is given by:

$$E_c = E_{GB} - E_s \quad (6)$$

To identify dislocations, we use the dislocation analysis tool (DXA) [76] integrated into the OVITO software [77], setting the trial Burgers circuit length to 8 atom-to-atom steps and the Burgers vector circuit stretchability to 9.

## 3. Discussions

### 3.1. Tilt and twist characters

#### 3.1.1. Energetic properties

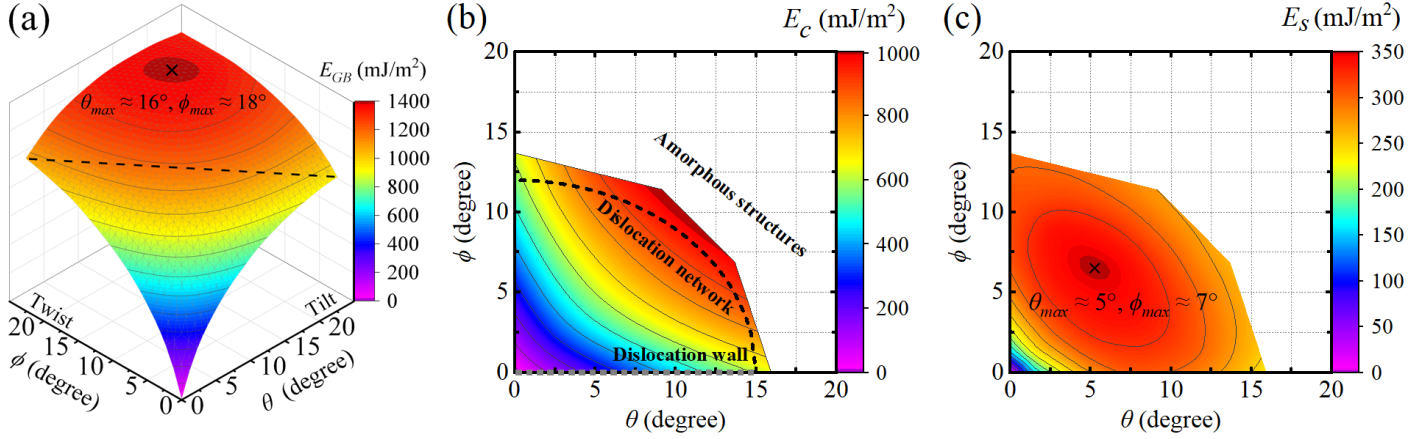

**Figure 2.** Energies of Si (001)/[100] mixed GB as the function of $\theta$ and $\phi$ in the examined two-dimensional GB character space. (a) $E_{GB}$; (b) $E_s$; (c) $E_c$. The black dash line in (a) denotes the mixed GBs dominated by the tilt characters have lower energies than those dominated by the twist characters. The colored area in (b) and (c) denotes that the mixed GBs there are dislocation network structures. The blank area in (b) and (c) denotes that the mixed GBs there are amorphous structures without identifiable dislocations. The grey dash line in (b) denotes that the symmetric tilt GBs at $\phi = 0°$ are dislocation arrays.

The variations of $E_{GB}$, $E_s$ and $E_c$ are given in Figure 2 as the function of $\theta$ and $\phi$. An unexpected $E_{GB}$ peak is shown in the position $\theta_{max} \approx 16°$ and $\phi_{max} \approx 18°$, as well as an expected misorientation-independent $E_{GB}$ that falls in a tilt angle range $0° \leq \theta \leq 16°$ and a twist angle range $0° \leq \phi \leq 18°$. Within this special area, $E_{GB}$ can be expressed as a function of both $\theta$ and $\phi$ but no longer depends on $\theta$ and $\phi$ if outside this area. The black dash line in Figure 2a also shows that the mixed GBs dominated by the tilt characters have lower energies than those dominated by the twist characters. In other words, this means that the mixed GBs which $\theta$ is greater than $\phi$ are energetically favorable.

Inspections of dislocation structures suggest that only 38 of the 121 examined (001)/[100] mixed GBs are dislocation network structures while the other 83 mixed GBs are amorphous structures without identifiable dislocations. The fact is also reflected by the blank area in Figures 2b and 2c where the computed dislocation energies $E_s$ and $E_c$ of 51 SAGBs (38 SAMGBs of total 121 (001)/[100] mixed GBs, seven small angle symmetric tilt GBs of total 11 (001)/[100] symmetric tilt GBs and six small angle twist GBs of total 11 (001) twist GBs) are plotted. The divisions between the colored and blank areas in Figures 2b and 2c highlight the irregular structural transitions from small angles (dislocation structures) to large angles (amorphous structures) in the two-dimensional GB character space. Despite the limitations in GB character sampling, the authors believe the slight differences and mutual perturbation between different dislocation cores are responsible for that. For example, the dislocation types of Si (001)/[100] small angle symmetric tilt and (001) small angle twist GBs are $b=\frac{1}{2}<110>$ mixed and $b=\frac{1}{2}<110>$ screw, respectively (see Figure 3 in *3.1.3. Dislocation structures*). The two types of $b=\frac{1}{2}<110>$ dislocations have different dislocation core radius $r_c$. For the corresponding GBs, small angle cutoffs $\theta_{small}$ and $\phi_{small}$ that maintain identifiable dislocation structures can be defined via Frank's formula following:

$$\theta_{small} \approx 2\arcsin\left(\frac{|b^{mixed}|}{2r_c^{mixed}}\right) \tag{7}$$

$$\phi_{small} \approx 2\arcsin\left(\frac{|b^{screw}|}{2r_c^{screw}}\right) \tag{8}$$

Where superscripts *mixed* and *screw* denotes the dislocation type. Equations (7) and (8) only provide small angle ranges of the symmetric tilt and twist GBs in the two-dimensional GB character space, while the perturbation between different co-existing dislocation cores in the mixed GB is still difficult to describe. To characterize the appearance of structural transitions from dislocation to amorphous structures, an elliptic equation is fitted by utilizing the positions of the 51 SAGBs in the two-dimensional GB character space following:

$$\frac{\theta^2}{4\arcsin^2\left(\frac{|b^{mixed}|}{2r_c^{mixed}}\right)} + \frac{\phi^2}{4\arcsin^2\left(\frac{|b^{screw}|}{2r_c^{screw}}\right)} \leq 1 \tag{9}$$

By using the precise values ($\theta_{small}$ = 15° [56] and $\phi_{small}$ = 12° [50]) determined in our previous work, equation (9) is plotted in Figure 2b as the black dash line, which matches well with the appearance of structural transitions in the two-dimensional GB character space. Moreover, another parameter that could be derived from equation (9) is the proportion of (001)/[100] SAMGBs, which is approximately 7% (estimated using the elliptic area).

Contrary to the expected rising trend of $E_c$, the variations of $E_s$ in Figure 2c shows a similar trend with $E_{GB}$. An $E_s$ peak is observed at the position $\theta_{max} \approx 5°$ and $\phi_{max} \approx 7°$, which is quite lower than the maximum $E_{GB}$ peak in the two-dimensional GB character space. Moreover, $E_s$ decreases rapidly for relatively higher misorientation angles, indicating that the increase of dislocation density is not always favorable for the population of dislocations stress near the SAGB.

*3.1.2. Dislocation structures*

Dislocation network structures of the 38 (001)/[100] SAMGBs extracted from the DXA tool are investigated, all of them contain infinite mixed dislocation arrays and discrete screw dislocation segments separated by the mixed dislocations. According to the relative relations between $\theta$ and $\phi$, the 38 SAMGBs could be generally divided into three types with similar topology and structural signatures: (a) $\phi$ is higher than $\theta$; (b) $\theta$ is equal to $\phi$; (c) $\theta$ is higher than $\phi$. Three dislocation network structures representing the corresponding SAMGB types are given in Figures 3a1, 3b1 and 3c1, respectively. 14 of the 38 SAMGBs are similar to the one in Figure 3a1, where screw dislocation junctions are observed between adjacent mixed dislocations. Only five SAMGBs are similar to Figure 3b1. The other 19 are similar to the one in Figure 3c1. The tilt character plays the dominant role and pure mixed dislocations appear. This SAMGB type has the highest proportion and relatively low energy among the three types, which means it should have a higher probability of occurrence.

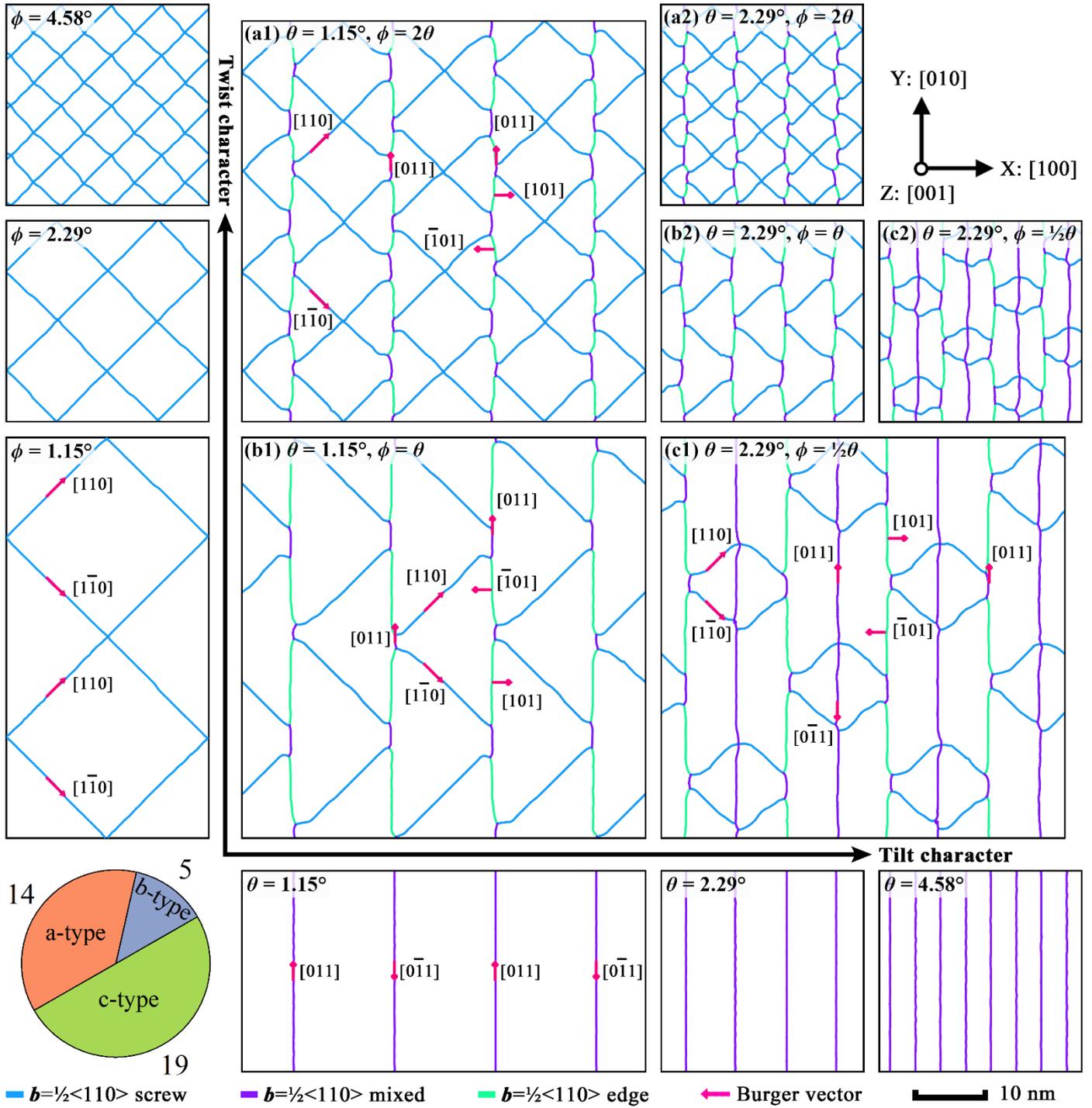

**Figure 3.** Dislocation structures of six Si (001)/[100] SAMGB characters and their tilt/twist components: (a1) $\theta$ = 1.15°, $\phi$ = 2.29°, TTR = 2, $A_t$ = 2.56°; (a2) $\theta$ = 2.29°, $\phi$ = 4.58°, TTR = 2, $A_t$ = 5.12°; (b1) $\theta$ = $\phi$ = 1.15°, TTR = 1, $A_t$ = 1.62°; (b2) $\theta$ = $\phi$ = 2.29°, TTR = 1, $A_t$ = 3.24°; (c1) $\theta$ = 2.29°, $\phi$ = 1.15°, TTR = 0.5, $A_t$ = 2.56°; (c2) $\theta$ = 4.58°, $\phi$ = 2.29°, TTR = 0.5, $A_t$ = 5.12°. The pie chart shows the amounts of SAMGBs from the three types (a), (b) and (c).

The dislocation network topology depends on the ratio between $\theta$ and $\phi$, which is easily explained by comparing Figures 3a1 to 3a2, 3b1 to 3b2 and 3c1 to 3c2 (the ratios between $\theta$ and $\phi$ are fixed to the same for both subfigures). This experimentally confirmed [49] conclusion is very useful as it allows explorations of almost arbitrary SAMGBs under acceptable simulation scales. To quantify the tilt and twist contributions from the two misorientation axis, we define the ratio between $\theta$ and $\phi$ (tilt-twist ratio, TTR) as a significant structural parameter for SAMGBs following:

$$TTR = \frac{\theta}{\phi} \qquad (10)$$

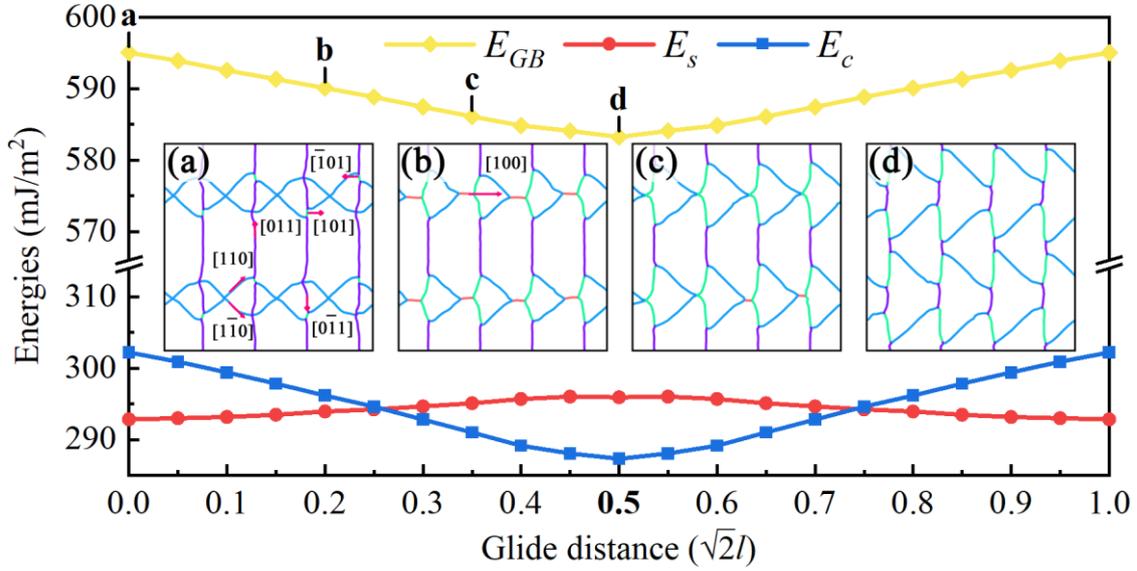

**Figure 4.** Energies of stable and metastable $\theta = \phi = 2.29°$, $TTR = 1$, $A_t = 3.24°$ Si (001)/[100] SAMGB phases versus glide distance of $b=½<110>$ screw dislocations. (a), (b) and (c) are metastable SAMGB phases. (d) is the stable SAMGB phase with the lowest excess energy. The red dislocation lines denote $b=<100>$ screw dislocations. The unit of glide distance is $\sqrt{2}l$. $l$ is the screw dislocation spacing determined by Frank's formula $l = |b|/2/\sin(½\phi)$. The captions for (a), (b), (c) and (d) follow the same in Figure 3.

Vdovin et al. [69] explained the formation mechanisms of Si (001)/[110] SAMGB structures as the glide of $b=½<110>$ screw dislocation, which mutually shifts half of the dislocation spacing and changes the Burger vector of the nearby mixed dislocations, but there was no exact evidence about the dislocation generation, glide and reaction as their observations only addressed stable dislocation structures that were well-developed. Here we use the SAMGB in Figure 3b2 as an example to clarify the details of its formation. As shown in Figures 4a-4d, the glide of $b=½<110>$ screw dislocation between adjacent mixed dislocations is observed by examining the metastable SAMGB phases. During the gliding process, $b=<100>$ screw and $b=½<110>$ edge dislocations generate, while $b=½[0\bar{1}1]$ mixed dislocation vanishes. The generated $b=<100>$ screw dislocation segment is from the excess proximity of two $b=½<110>$ screw dislocation segments, and the generated edge dislocation has two Burger vectors ($b=½[\bar{1}01]$ and $b=½[101]$) from two dislocation reactions, respectively as:

$$½[011] \rightarrow ½[110] + ½[\bar{1}01] \qquad (11)$$

$$½[1\bar{1}0] + ½[011] \rightarrow ½[101] \qquad (12)$$

The SAMGB energies versus the glide distance of $b=½<110>$ screw dislocation are plotted in the diagram of Figure 4, from which the dislocation reactions gradually decrease $E_s$ but slightly increase $E_c$. From 0 to 0.5 unit glide distance, $E_{GB}$ decreases and the most stable phase is located at 0.5 unit glide distance, both of which indicate that the glide is driven by the energy release of dislocation reactions. These metastable SAMGB phases also show stability against the quenching process. As the time scale of the MD simulation is lower than the experiments for several orders of magnitude, the dislocation glide to form stable SAMGB phases may be processed in macroscopic relaxation at a longer time scale. The authors are encouraged by the observation of various metastable SAMGB phases, which not only becomes evidence for their formation mechanisms but also provides new insights into possible SAMGB phase transformations.

## 3.2. Rotation character

### 3.2.1. Dislocation properties

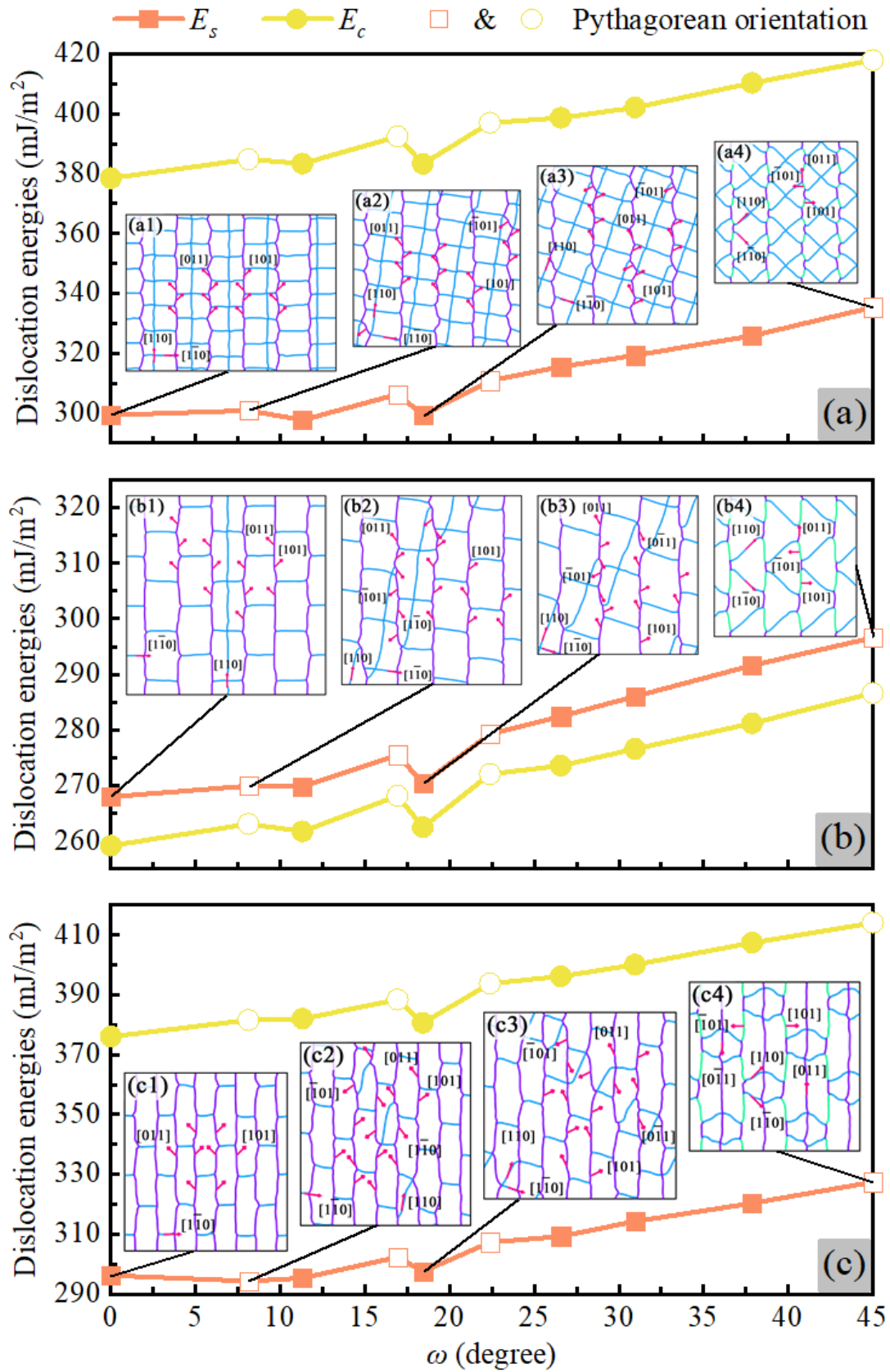

**Figure 5.** Dislocation structures and energies of three given (001) SAMGB characters as the function of $\omega$: (a) The SAMGB character shown in Figure 3a2; (a1) $\theta = 2.43°$, $\phi = 4.58°$, $TTR = 1.88$, $A_t = 5.18°$ (001)/[110] SAMGB; (a2) $\theta = 2.29°$, $\phi =$

4.58°, $TTR = 2$, $A_t = 5.12°$ (001)/[430] SAMGB; (a3) $\theta = 2.05°$, $\phi = 4.58°$, $TTR = 2.23$, $A_t = 5.02°$ (001)/[210] SAMGB; (a4) $\theta = 2.29°$, $\phi = 4.58°$, $TTR = 2$, $A_t = 5.12°$ (001)/[100] SAMGB; (b) The SAMGB character shown in Figure 3b2; (b1) $\theta = 2.43°$, $\phi = 2.29°$, $TTR = 1.06$, $A_t = 3.34°$ (001)/[110] SAMGB; (b2) $\theta = \phi = 2.29°$, $TTR = 1$, $A_t = 3.24°$ (001)/[430] SAMGB; (b3) $\theta = 2.05°$, $\phi = 2.29°$, $TTR = 0.90$, $A_t = 3.07°$ (001)/[210] SAMGB; (b4) $\theta = \phi = 2.29°$, $TTR = 1$, $A_t = 3.24°$ (001)/[100] SAMGB; (c) The SAMGB character shown in Figure 3c2; (c1) $\theta = 4.86°$, $\phi = 2.29°$, $TTR = 0.47$, $A_t = 5.37°$ (001)/[110] SAMGB; (c2) $\theta = 4.58°$, $\phi = 2.29°$, $TTR = 0.5$, $A_t = 5.12°$ (001)/[430] SAMGB; (c3) $\theta = 4.10°$, $\phi = 2.29°$, $TTR = 0.56$, $A_t = 4.69°$ (001)/[210] SAMGB; (c4) $\theta = 4.58°$, $\phi = 2.29°$, $TTR = 0.5$, $A_t = 5.12°$ (001)/[100] SAMGB. The Pythagorean orientations are marked specially, which ensures that $\theta$, $\phi$ and $TTR$ are fixed during the rotation. $\theta$, $\phi$ and $TTR$ of the other lateral orientations vary a little from the initially given mixed GB characters due to the sampling limitations. Table A2 in the Appendix shows the details of all examined SAMGBs in Figure 5. The captions for (a), (b), (c) and (d) follow the same in Figure 3.

A rotation character is implicitly contained in the mixed GB character. In Figure 3, the screw dislocation network has rotated 45° relative to the mixed dislocation array. This rotation angle $\omega$ is determined by the lateral orientations $i$ and $j$ of the GB plane without additional effects on the normal of the GB plane. To investigate the rotation character, $\theta$ and $\phi$ are better fixed to constants, and thus only the Pythagorean numbers are available to sample. Alternatively, the authors try to sample other non-Pythagorean orientations with minimum variations to the original $\theta$ and $\phi$. In that case, the three mixed GB characters in Figure 3 are used as examples to study how the rotation character affects the SAMGB properties.

Figure 5 shows the computed dislocation structures and energies of the three (001) mixed GB characters as the function of $\omega$ (noting that the physically significant rotations are in $0° \leq \omega \leq 45°$ as rotations $+\omega$ and $-\omega$ are crystallographic equivalents for the <100> axis). In general, for the three mixed GB characters, the minimum dislocation energies are always obtained at $\omega = 0°$ where the lateral orientation is the <110> axis. Both $E_s$ and $E_c$ increase with $\omega$, but show energy valleys around $\omega = 12°$ and $\omega = 18°$. These energy valleys are generated from the inaccuracy of sampling (i.e., $\theta$ and $\phi$ are not fixed during the rotation), except for this, the general trends of $E_s$ and $E_c$ of the three mixed GB characters seem to have close-linear dependences with $\omega$. It should be noticed that the authors do not intend to examine the dislocation energies span the three-dimensional GB character space ($\theta$, $\phi$ and $\omega$) because accurately sampling these variables in acceptable simulation scales is almost impossible, and therefore such assumption about the dependences of dislocation energies has its uses and will be discussed later.

For the dislocation structures in Figures 5a1-5a4, 5b1-5b4 and 5c1-5c4, no substantial changes have been made to the previously discussed formation mechanisms. The noteworthy point is that the infinite $\boldsymbol{b}$=[101] and $\boldsymbol{b}$=[011] mixed dislocations substitute $\boldsymbol{b}$=½[110] screw dislocations via dislocation reaction following:

$$\tfrac{1}{2}[011] + \tfrac{1}{2}[0\bar{1}1] + \tfrac{1}{2}[110] \rightarrow \tfrac{1}{2}[101] + \tfrac{1}{2}[011] \tag{13}$$

Equation (13) is characterized in Figure 6a and is invalid for (001)/[100] SAMGBs at $\omega = 45°$. At first sight, the additional infinite $\boldsymbol{b}$=½[110] screw dislocation is unreasonable in Figure 6b1. This is because the spacings of mixed dislocation and screw dislocation do not satisfy the designated ratio. According to Frank's formula, the infinite screw dislocation can be eliminated when $\theta$ is equal to $\sqrt{2}\,\phi$ for the (001)/[110] SAMGB character.

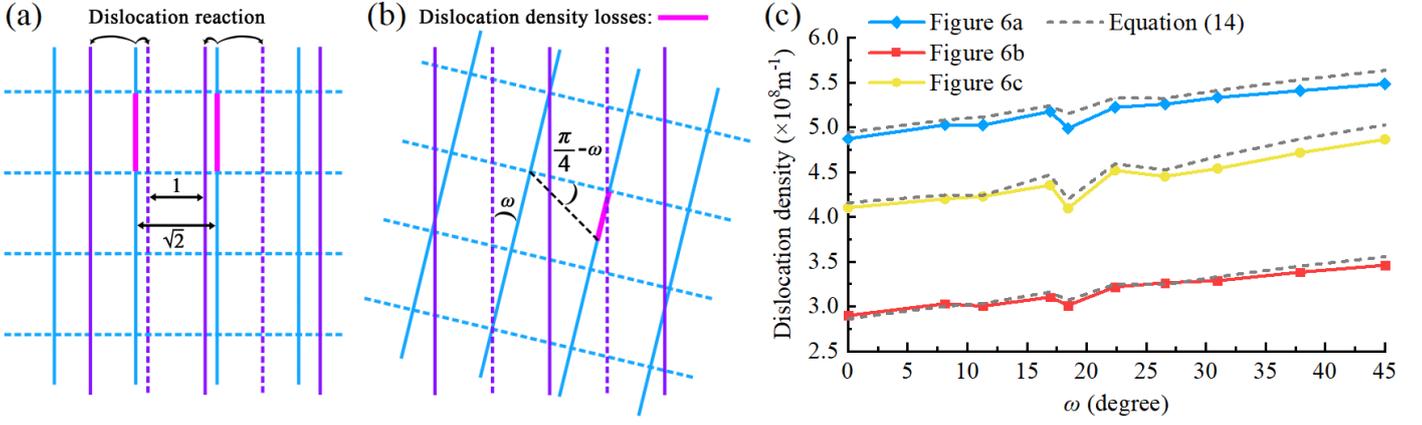

**Figure 6.** (a) Dislocation reaction mechanisms for (001)/[110] SAMGB character ($\theta = \phi$); (b) Dislocation density losses during the rotation, which is the supplementary of SAMGB formation in Figure 4 at $0° \leq \omega < 45°$; (c) Simulated dislocation density against $\omega$, along with the grey reference values from equation (15). For (a) and (b), the purple line is $\boldsymbol{b}$=½[011] mixed dislocation, the purple dash line is $\boldsymbol{b}$=½[01$\bar{1}$] mixed dislocation, the blue line is $\boldsymbol{b}$=½[110] screw dislocation and the blue dash line is $\boldsymbol{b}$=½[1$\bar{1}$0] screw dislocation.

The correlations between dislocation density and $\omega$ are derived considering the dislocation reaction of equation (13). Take the (001)/[110] SAMGB character for an example, its dislocation density depends on the minimum between the mixed dislocation density and screw dislocation density. The logic is simple: if the spacing of screw dislocation is little than the mixed dislocation, then the mixed dislocations substitute the same amount of screw dislocations; if the spacing of screw dislocation is greater than the mixed dislocation, then all screw dislocations are substituted by the mixed dislocations. Except this, the other parts of the dislocation network are simplified to the overlap of the tilt and twist components as the glide of screw dislocations causes no extra dislocation density losses. Moreover, from Figures 5b2 and 5b3, the dislocation reaction of equation (13) also results in the variations of SAMGB formation because the $\boldsymbol{b}$=½[110] screw dislocations are partially substituted by the mixed dislocations at $0° < \omega < 45°$, which is illustrated in Figure 6b2 as a function of $\tan(\pi/4 - \omega)$. Therefore, the dislocation density $D$ for an arbitrary (001) SAMGB is derived as a function of $\theta$, $\phi$ and $\omega$ following:

$$D(\theta,\phi,\omega) = \frac{4\sin\left(\frac{\theta}{2}\right)}{\sqrt{2}\left|\boldsymbol{b}^{mixed}\right|} + \frac{4\sin\left(\frac{\phi}{2}\right)}{\left|\boldsymbol{b}^{screw}\right|} - \tan\left(\frac{\pi}{4} - \omega\right) \times \min\left[\frac{2\sin\left(\frac{\theta}{2}\right)}{\sqrt{2}\left|\boldsymbol{b}^{mixed}\right|}, \frac{2\sin\left(\frac{\phi}{2}\right)}{\left|\boldsymbol{b}^{screw}\right|}\right] \quad (14)$$

The simulated dislocation density is plotted in Figure 6c for verification. The reference values from equation (14) agree well with the simulated values. Within the framework of equation (14), the rotation character can be regarded with a close-linear impact on the dislocation properties, which will benefit the subsequent work.

### 3.2.2. A revised Read-Shockley relationship

For the SAMGB energies, classical descriptions like the Read-Shockley relationship are desired as SAMGB structures can be characterized as dislocation models. However, the three co-existing DOFs in the mixed GB character make the one-DOF-based classical Read-Shockley relationship unsuitable to describe the GB energy surface therein. But at the same time, we can still see the strong universality of the classical Read-Shockley relationships due to the general trend of GB energy surface. Hence, we consider revising the classical

Read-Shockley relationship for the mixed GB character. In their most general cases, the Read-Shockley relationships for small angle symmetric tilt and twist GBs are respectively defined as the following:

$$E_{GB}^{tilt}(\theta) = \frac{\theta\left[E_c^{u,tilt} - E_s^{u,tilt}\ln(\theta)\right]}{\left|\boldsymbol{b}^{tilt}\right|} \tag{15}$$

$$E_{GB}^{twist}(\phi) = \frac{\phi\left[E_c^{u,twist} - E_s^{u,twist}\ln(\phi)\right]}{\left|\boldsymbol{b}^{twist}\right|} \tag{16}$$

Where superscripts *mixed*, *tilt* and *twist* denote the GB type. Constants $E_c^u$ and $E_s^u$ represent the dislocation core energy and strain energy per unit length, respectively. Their values are given in Table 2 by fitting the energies of 11 symmetric tilt GBs and 11 twist GBs as the function of $\theta$ or $\phi$, respectively.

**Table 2.** Dislocation energy parameters and RMSE values for the Read-Shockley relationships of pure small angle symmetric tilt and twist GBs.

| GB type | $E_c^u$ (eV/Å) | $E_s^u$ (eV/Å) | $R^2$ (mJ/m$^2$) | RMSE (mJ/m$^2$) |
|---|---|---|---|---|
| (001)/[100] symmetric tilt | 0.1295 | 0.7282 | 0.9782 | 5.52 |
| (001) twist | 0.1579 | 0.5296 | 0.9739 | 6.82 |

If possible, the Read-Shockley relationship for the mixed GB can be revised using the following equation:

$$E_{GB}^{mixed} = E_{GB}^{tilt}(\theta) + E_{GB}^{twist}(\phi) + \Delta E(\theta,\phi) \tag{17}$$

Where $\Delta E$ is the energy variations when the tilt and twist components form the mixed tilt-twist character. Figure A2 in the appendix shows $\Delta E$ as a function of $\theta$ and $\phi$, and our optimum fitting analysis suggests that it must satisfy the following form (or any similar form if applicable)

$$\Delta E(\theta,\phi) = E_0 \frac{\theta\phi}{(\theta+\alpha)(\phi+\beta)} \tag{18}$$

Where $\alpha = 0.2969$rad, $\beta = 0.4562$rad and $E_0 = -2570.84$mJ/m$^2$ are fitting parameters. $R^2$ of this fitting is 0.9611. The special form of equation (18) ensures equation (17) degenerates to the original Read-Shockley relationship at $\theta = 0$ or $\phi = 0$. Thus, the revised Read-Shockley relationship for Si (001)/[100] mixed GBs could be written as:

$$E_{GB}^{mixed}(\theta,\phi) = E_{GB}^{tilt}(\theta) + E_{GB}^{twist}(\phi) + E_0 \frac{\theta\phi}{(\theta+\alpha)(\phi+\beta)} \tag{19}$$

Going further, for arbitrary SAMGBs located on the (001) plane, we assume the rotation character has a close-linear impact on the SAMGB energies based on the previous discussions, then equation (19) can be improved by taking account of $\omega$ following

$$E_{SAMGB}^{mixed}(\theta,\phi,\omega) = E_{GB}^{tilt}(\theta) + E_{GB}^{twist}(\phi) + E_0 \frac{\theta\phi}{(\theta+\alpha)(\phi+\beta)}\left(1+k\times\tan\left(\frac{\pi}{4}-\omega\right)\right) \tag{20}$$

Where $k = 1.7203$ is a fitting parameter. $R^2$ of this fitting is 0.9794. Note that equation (19) is for the studied Si (001)/[100] mixed GBs at $\omega = \pi/4$ while equation (20) is for the Si (001) SAMGBs. Examinations of their performance are given in Figure 7.

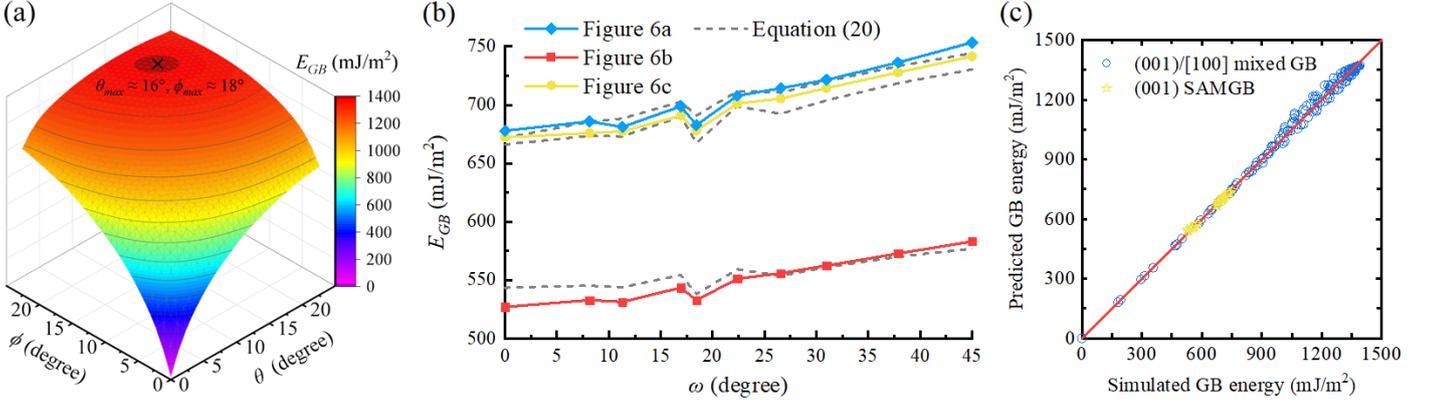

**Figure 7.** (a) Predicted energy surface of Si (001)/[100] mixed GB calculated from equation (19) in the examined two-dimensional GB character space; (b) Simulated energies of the three given SAMGB characters against $\omega$, along with the grey reference values from equation (20); (c) Predicted GB energies from equations (19) and (20) against simulated GB energy. The blue circle denotes the data of 121 (001)/[100] mixed GB characters in Figure 7a, and the yellow five-pointed star denotes the data of 30 (001) SAMGB characters in Figure 7b.

As shown in Figure 7a, equation (19) provides relatively good predictions and a RMSE at around 21.63mJ/m² about the GB energy surface. Even for the energies of those amorphous GB structures, its extrapolation ability is also robust as it correctly captures the energy peak. The prediction from equation (20) also matches well with the simulated SAMGB energies in Figure 7b, along with a low RMSE at around 11.52mJ/m². The authors are optimistic that these equations and their fitting procedure will be useful for learning SAMGB structure-property relationships spanning the five-dimensional GB character space if future work considers the role of the GB plane normal.

## 3.3. Validation and Comparison

### 3.3.1. Theoretical calculations

To confirm the validity of the SAMGB energies extracted from atomistic simulations, we adopt the SAMGB energy model of Zhang et al. [66], which is based on the Continuum framework of Zhu and Xiang [78]. The framework introduced a dislocation density potential function $\eta$ derived from the FBE, and the interfacial dislocation structures can be described in terms of $\nabla\eta$. In the SAMGB energy model, a given SAMGB structure is approximated by a series of pure and infinite dislocation arrays, no matter whether it contains discrete dislocation segments or not. Under that case, the SAMGB energy $E_{SAMGB}$ comprised of $i$ sets of dislocation arrays is written as:

$$E_{SAMGB} = \sum_i \frac{\mu|\boldsymbol{b}_i|^2}{4\pi(1-\nu)}\left[1-\nu\frac{(\nabla\eta_i \times \boldsymbol{n}\cdot\boldsymbol{b}_i)^2}{|\boldsymbol{b}_i|^2\|\nabla\eta_i\|^2}\right]\cdot\|\nabla\eta_i\|\log\left(\frac{1}{r_g\sqrt{\|\nabla\eta_i\|^2+\varepsilon}}\right) \quad (21)$$

Where $\mu$ is the shear modulus, $\nu$ is the Poisson ratio, $\boldsymbol{n}$ is the normal of the GB plane, and $\boldsymbol{b}_i$ is the Burger vector of dislocation array $i$. $r_g$ and $\varepsilon$ are empirical parameters related to the dislocation core radius and dislocation core energy, respectively. According to the Continuum framework, the calculation of $\nabla\eta_i$ for dislocation array $i$ must follow the stable SAMGB structures to ensure its accuracy. Furthermore, the selection of $r_g$ and $\varepsilon$ would greatly affect the results. We use the recommended values ($r_g = \boldsymbol{b}$ and $\varepsilon = 8\times10^{-7}\boldsymbol{b}^{-2}$) of reference [66] and constant temperature elastic properties ($\mu = 66.8$GPa, $\nu = 0.31$) of silicon while utilizing equation (21). Here we give two examples in Figure A3 in the Appendix to show how the simulated structures input into the SAMGB energy model for energy calculations.

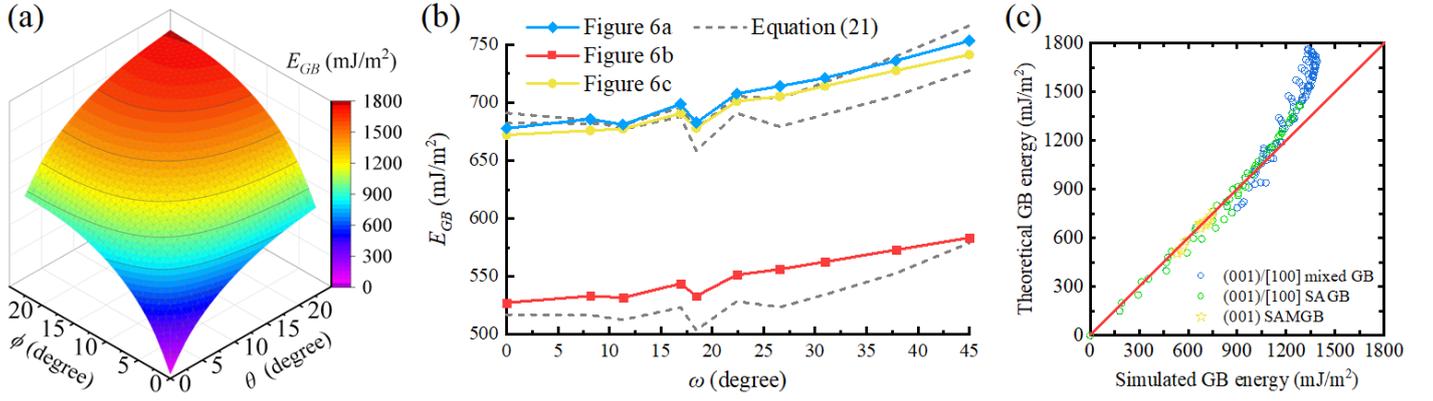

**Figure 8.** (a) Theoretical GB energy calculated from equation (21) as a function of $\theta$ and $\phi$ in the two-dimensional GB character space. (b) Simulated energies of the three given SAMGB characters against $\omega$, along with the grey reference values from equation (21); (c) Theoretical GB energy calculated from equation (21) against simulated GB energy. The 51 (001)/[100] SAGBs are colored with green, and the other 83 mixed GBs without identifiable dislocation structures are colored with blue. The 30 (001) SAMGB energies in Figure 8b are marked with the yellow five-pointed star.

The theoretical GB energy calculated from equation (21) is given in Figure 8a as the function of $\theta$ and $\phi$, and no energy peaks are observed compared with Figure 2a. For the 83 amorphous GB structures, equation (21) gives relatively higher energy than simulations. On the contrary, the theoretical GB energy is slightly lower than the simulation for the rotation character in Figure 8b, and both suggest the same trend of energy variations against $\omega$. Figure 8c plots the theoretical energy against the simulated energy. The atomistic simulation actually reaches agreements with the SAMGB energy model for the energies of all examined SAMGBs, the RMSE is 34.03mJ/m². There are several reasons why the simulation slightly deviates from this theoretical model. First, the overlap of dislocation cores may be responsible for the low simulated GB energies of amorphous GB structures (shown as blue circles in Figure 8c). Second, the SAMGB energy model uses approximation for discrete dislocation segments and therefore generates deviations.

In comparison between the atomistic simulation and theoretical model, the authors wish to highlight the significance of their revised Read-Shockley relationships. On the one hand, the SAMGB energy model is able to solve the energies of arbitrary known SAMGB structures, but its approximation into pure and infinite dislocation arrays has limitations due to the dislocation glide and reaction. In some cases (e.g., the rotation character), the SAMGB structures are completely unknown, and thus further simulation/experiment is desired. On the other hand, although the revised Read-Shockley relationships lack transferability, they have better performances than the SAMGB energy model on the specific focus. The relationships allow one to directly calculate the SAMGB energies via three macroscopic structural descriptors ($\theta$, $\phi$ and $\omega$) without pre-knowing its dislocation structures. Broadly speaking, the roles of these equations are not conflict but mutually complementary.

*3.3.2. Experimental observations*

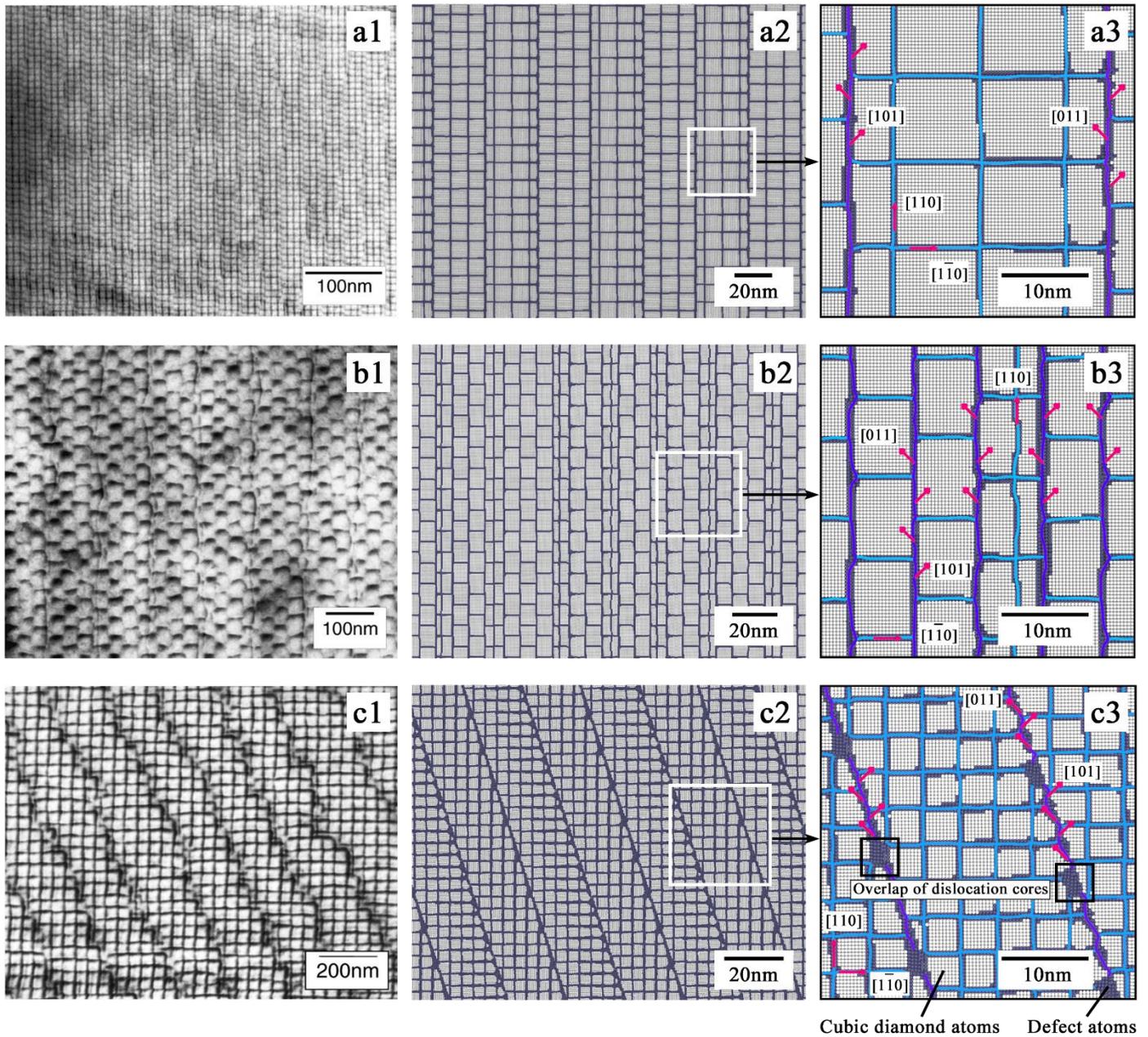

**Figure 9.** Experimental observations [49] and atomistic simulations of three Si (001) SAMGBs: (a1) TEM image of $\theta = 0.54 \pm 0.04°$, $\phi = 2.55 \pm 0.05°$, $TTR \approx 4.72$ SAMGB; (a2) Simulated atomic structures of $\theta = 0.54°$, $\phi = 2.31°$, $TTR = 4.24$, $A_t = 2.37°$ SAMGB; (a3) Dislocation structures of a part from (a2), marked by the white rectangle; (b1) TEM image of $\theta = 0.53 \pm 0.04°$, $\phi = 0.69 \pm 0.02°$, $TTR \approx 1.30$, $A_t = 0.84 \pm 0.03°$ SAMGB; (b2) Simulated atomic structures of $\theta = 2.22°$, $\phi = 2.44°$, $TTR = 1.10$, $A_t = 3.29°$ SAMGB, the experimental misorientation angles are amplified by 4 times; (b3) Dislocation structures of a part from (b2), marked by the white rectangle; (c1) TEM image of $\theta = 0.083°$, $\phi = 0.54°$, $TTR \approx 6.49$, $A_t = 0.55°$ SAMGB; (c2) Simulated atomic structures of $\theta = 0.81°$, $\phi = 5.45°$, $TTR = 6.71$, $A_t = 5.51°$ SAMGB, the experimental misorientation angles are amplified by 10 times; (c3) Dislocation structures of a part from (c2), marked by the white rectangle; (a1), (b1) and (c1) are reprinted with the permission from *Springer Nature*.

To confirm the reliability of the atomistic simulation, the simulation procedure is used to reproduce the observed SAMGB structures in existing references. To satisfy an acceptable simulation scale, the author's strategy is to set the parameter *TTR* of the simulations close to the experiments and compare the dislocation network topology under the conclusion that the topology only varies with *TTR*.

The comparisons between the simulations and the observations of Akatsu et al. [49] are shown in Figure 9. The annealing temperature of each simulation is strictly following the experimental value [49]. Figures 9a1 and 9b1 show the TEM images of two Si (001)/[110] SAMGBs, which are in excellent agreement with the subsequent simulations. The screw dislocation network is separated by the infinite mixed dislocations and mutually shifted half of their spacing. There are also additional infinite screw dislocations in Figure 9b1 when $\phi$ is a little higher than $\theta$, and thus confirms the validity of Figures 5b1-5b4. The SAMGB in Figure 9c1 is approximated by a (001)/[210] SAMGB in the simulation of Figure 9c2. The intersections of mixed and screw dislocations become a defect cluster without identifiable dislocations in Figure 9c3, experiments also show this detail as there are extremely dark contrasts near the mixed dislocation array.

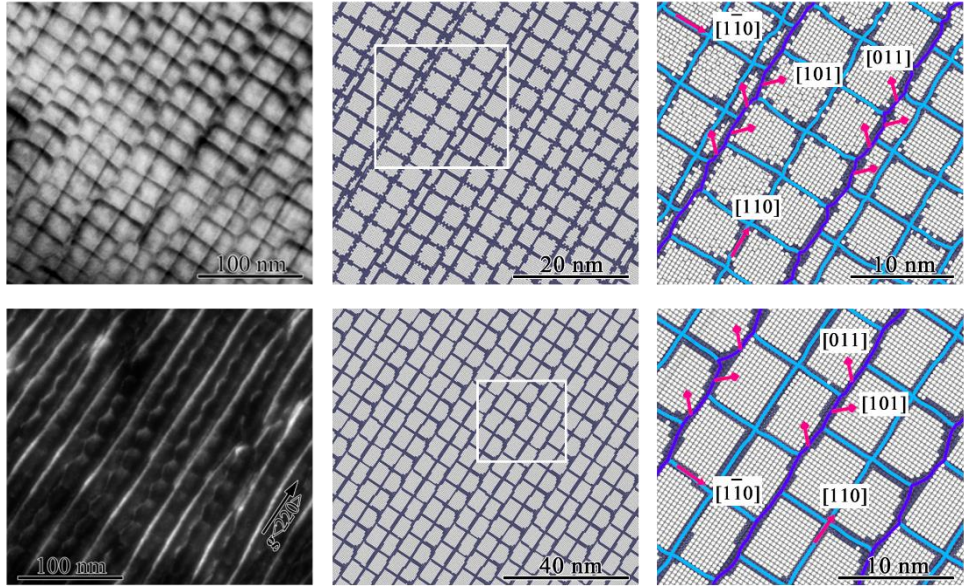

**Figure 10.** Experimental observations [69] and atomistic simulations of two Si (001)/[110] SAMGBs: (a1) TEM image of a (001)/[110] SAMGB, its $\theta$ and $\phi$ are not given in [69]; (a2) Simulated atomic structures of $\theta = 1.39°$, $\phi = 4.58°$, *TTR* = 3.29, $A_t = 4.79°$ SAMGB; (a3) Dislocation structures of a part from (a2), marked by the white rectangle; (b1) TEM image of a (001)/[110] SAMGB, its $\theta$ and $\phi$ are not given in [69]; (b2) Simulated atomic structures of $\theta = 1.60°$, $\phi = 3.40°$, *TTR* = 2.13, $A_t = 3.76°$ SAMGB; (b3) Dislocation structures of a part from (b2), marked by the white rectangle. (a1) and (b1) are reprinted with the permission from *Trans Tech Publications*.

We then use the revealed formation mechanisms of SAMGBs to reproduce the observations of Vdovin, et al. [69], in which the precise misorientation angles are not given. As shown in Figure 10, the atomistic simulation successfully approximates the two SAMGB structures with the aid of the formation mechanisms and without the prerequisite that knowing the accurate misorientation angles. Although the simulation procedure is capable of generally reproducing the macroscopic SAMGB structures, it still slightly deviates from the experiments. For example, some of the mixed dislocations are curved to a zigzag line, which is not well captured by the simulation. Furthermore, the formation mechanisms indicate the SAMGB structures should contain infinite mixed dislocation arrays in any case, which are difficult to explain the appearance of the long honeycomb dislocation network structures. The formation mechanisms of SAMGBs are therefore anticipated to vary with the GB plane normal.

# 4. Conclusions

In this work, the structures and energies of Si (001) SAMGBs with co-existing tilt and twist components are computed as the function of three macroscopic GB characters at the atomic scale. The relevant results are further validated by both theoretical calculations and experimental observations. The noteworthy conclusions are summarized below:

(1) GB energy, dislocation strain energy and dislocation core energy are plotted as a function of $\theta$ and $\phi$. The GB energy and dislocation strain energy respectively show energy peaks at $\theta \approx 16°$, $\phi \approx 18°$ and $\theta \approx 5°$, $\phi \approx 7°$, while the dislocation core energy increases with $\theta$ and $\phi$. From $\omega = 0°$ to $\omega = 45°$, the GB energy increases almost linearly, based on which a revised Read-Shockley relationship capable of precisely describing the GB energy variations is fitted as a function of $\theta$, $\phi$ and $\omega$.

(2) GB structural transitions from dislocation to amorphous structures in the two-dimensional GB character space are fitted as a function of $\theta$, $\phi$ and dislocation core radii, from which nearly 7% of the total mixed GB population is estimated to be SAMGBs that comprised of dislocation network structures, and they can be characterized and classified into three types by the ratio between $\theta$ and $\phi$. The proportion, topology and structural signatures of each type are given.

(3) Detailed dynamic process of SAMGB formation is extracted from the metastable phases, and the formation mechanisms are explained as the segmentation of mixed dislocation arrays and two energetically favorable dislocation reactions. One reaction causes the mutual glide of separated screw dislocations, and the other one results in the losses of screw dislocations, from which we derive the dislocation density of these SAMGBs as a function of $\theta$, $\phi$ and $\omega$.

(4) This work is validated and supported by both theoretical calculations and experimental observations. The simulated energies are compared with the Continuum energy model of SAMGB. Based on the consistency with this theoretical model, the significance of the revised Read-Shockley relationship is highlighted. The atomistic simulation is also used to reproduce the experimental observations of silicon (001) SAMGBs according to the conclusion that the dislocation network topology does not vary with the ratio between $\theta$ and $\phi$. Our simulation procedure is proved reliable as it successfully reproduces many macroscopic observations at the atomic scale.

Some of the conclusions may be unsurprising but reinforced many known facts about SAMGBs. For example, the formation mechanisms of well-developed (001)/[110] SAMGB structures and the role of the parameter *TTR* are unsurprising for the earlier experimental efforts [49, 69]. However, the authors intend to provide insight into SAMGBs from their correlations with the decomposed tilt and twist characters, so one can control the nano-patterns in bonded thin films via the revealed formation mechanisms. The examined (001)/[100] SAMGB structures are also of interest as previous reports seldom addressed this GB character. Most importantly, the energetic properties are computed and metastable SAMGB phases are obtained, which are inaccessible for experimental methods to collect. The SAMGB formation is therefore re-explained using the structures and energies of metastable phases. Moreover, the revised Read-Shockley relationship may be successful for the small angle portion, but the general trend of GB energy surface towards higher tilt and twist angles still needs to be further examined beyond this work. It is expected to find some low-energy ground-state structures in some special mixed GB characters.

Notably, this work only addresses a special type of mixed GB character, namely the mixed symmetric tilt and twist character, which geometry allows periodic conditions for modelling. The situation becomes complicated while studying the mixed asymmetric tilt and twist character. Future work should consider this to completely clarify the last puzzle in the five-dimensional GB character space.

## Data availability

The LAMMPS code and numerical data are available upon reasonable request.

## Acknowledgements

The authors thank Prof. W.N. Zou (Dean at Institute for Advanced Study, Nanchang University) for his support.

## Appendix

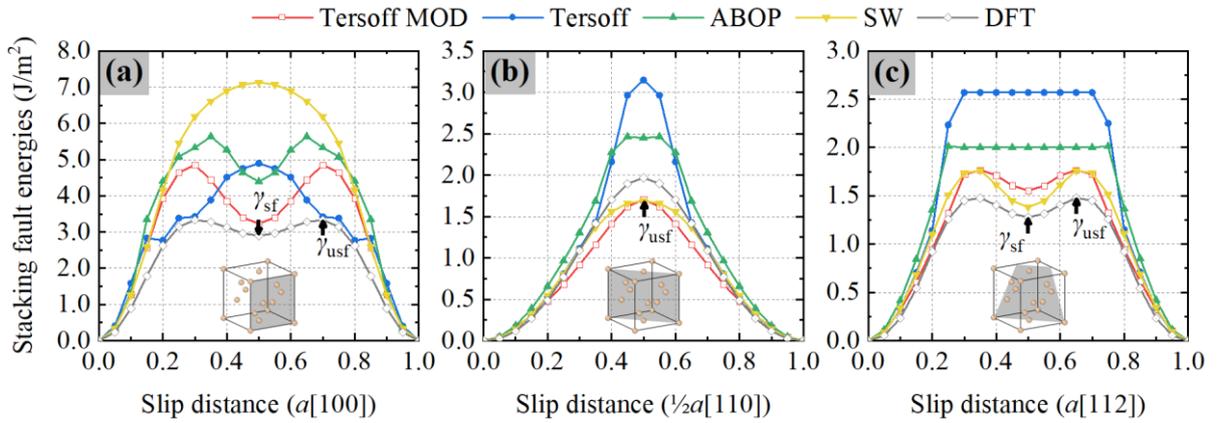

**Figure A1** Stacking fault energies of DFT and semi-empirical potentials obtained by VASP and LAMMPS: (a) [100] orientation; (b) [110] orientation; (c) [112] orientation. $\gamma_{sf}$ and $\gamma_{usf}$ represent stable and unstable stacking fault energies, which relate to the dislocation dissociation and nucleation, respectively.

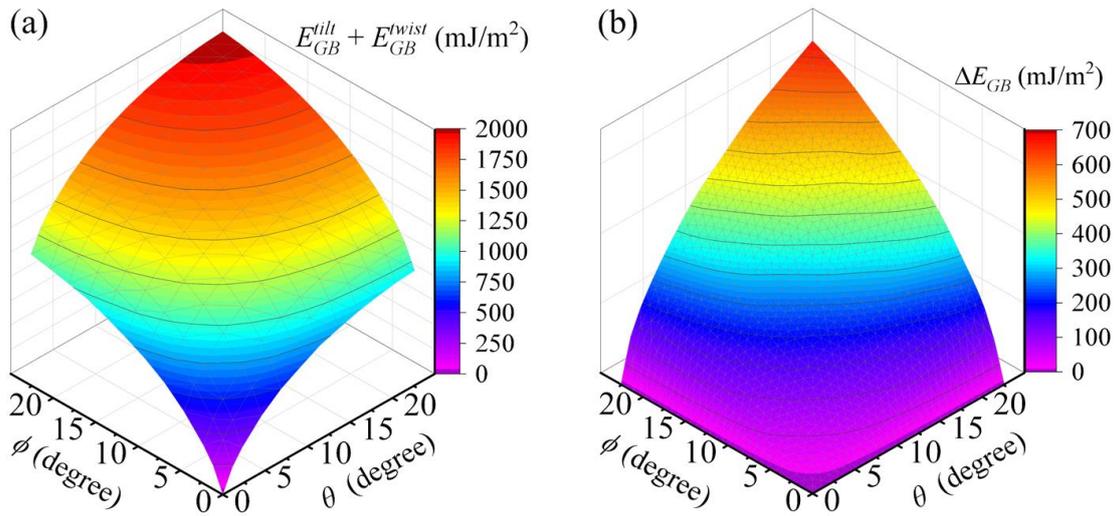

**Figure A2.** Parameters in the revised Read-Shockley relationships as the function of $\theta$ and $\phi$. (a) $E_{GB}^{tilt} + E_{GB}^{twist}$; (b) $\Delta E_{GB}$. (a) is the sum of the energies of the tilt components and the twist components calculated using the data from 11 symmetric tilt and 11 twist GBs. (b) is the variation of energy when the tilt and twist components form the mixed GB character, which is defined following $\Delta E_{GB} = E_{GB}^{tilt} + E_{GB}^{twist} - E_{GB}^{mixed}$.

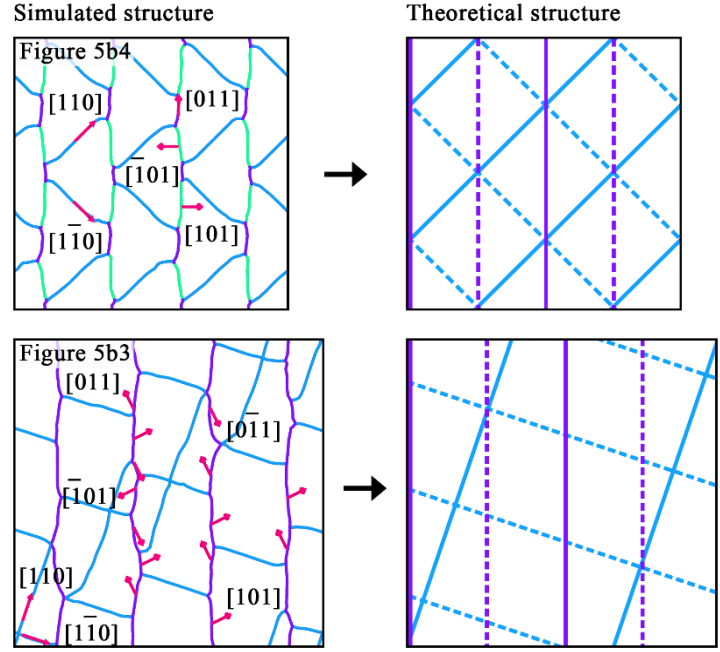

**Figure A3.** Two examples used in the SAMGB energy model. The SAMGBs generated from atomistic simulations contain discrete dislocation segments and the SAMGB energy model approximately treated them as pure and infinite dislocation arrays with the same total Burger vector length. In other words, they are considered as the overlap of their tilt and twist components, but the overlap position (applicable when the set of dislocation arrays is greater than 2) is determined by the constrained and unconstrained minimizations in the Continuum framework. The purple line is $b=½[011]$ mixed dislocation, the purple dash line is $b=½[0\bar{1}1]$ mixed dislocation, the blue line is $b=½[110]$ screw dislocation and the blue dash line is $b=½[1\bar{1}0]$ screw dislocation.

**Table A1** Physical properties of diamond silicon from experiments, DFT and semi-empirical potentials [57].

| Properties | Experiments | DFT | Semi-empirical potentials | | | |
|---|---|---|---|---|---|---|
| | | | Tersoff | **Tersoff MOD** | ABOP | SW |
| $a$ (Å) | 5.43 | 5.46 | 5.43 | **5.43** | 5.43 | 5.43 |
| $E_{coh}$ (eV) | -4.63 | -5.42 | -4.62 | **-4.63** | -4.63 | -4.34 |
| $E_v$ (eV) | 3.6±0.2 | 3.67 | 3.70 | **3.72** | 3.20 | 2.82 |
| $C_{11}$ (GPa) | 168 | 153 | 143 | **166** | 167 | 162 |
| $C_{12}$ (GPa) | 65 | 56 | 75 | **65** | 65 | 82 |
| $C_{44}$ (GPa) | 80 | 72 | 69 | **77** | 72 | 60 |
| $T_m$ (K) | 1687 | – | 2396[a] | **1681** | 2547 | 1688 |

$a$, $E_{coh}$, $E_v$, $C_{ij}$ and $T_m$ are the lattice constant, cohesive energy, vacancy formation energy, elastic constants and melting point, respectively.

**Table A2** Details of the examined SAMGB characters in Figure 7. θ, φ and TTR of SAMGBs 1–9 are set close to SAMGB 10 to ensure other variables are fixed to the same while examining dislocation structures and energies under different ω.

| ID | Order | θ (degree) | φ (degree) | TTR | $A_t$ (degree) | Lateral orientation | ω (degree) | X size (nm) | Y size (nm) |
|---|---|---|---|---|---|---|---|---|---|
| a | 1 | 2.43 | 4.58 | 1.88 | 5.18 | [110] | 0.00 | 76.9 | 19.2 |
| a | 2 | 2.29 | 4.58 | 2 | 5.12 | [430] | 8.13 | 135.9 | 67.9 |
| a | 3 | 2.22 | 4.58 | 2.06 | 5.09 | [320] | 11.31 | 196.0 | 49.0 |
| a | 4 | 2.29 | 4.58 | 2 | 5.12 | [15 8 0] | 16.93 | 231.0 | 115.5 |
| a | 5 | 2.05 | 4.58 | 2.23 | 5.02 | [210] | 18.43 | 30.4 | 30.4 |
| a | 6 | 2.29 | 4.58 | 2 | 5.12 | [12 5 0] | 22.38 | 176.7 | 88.3 |
| a | 7 | 2.17 | 4.58 | 2.11 | 5.07 | [310] | 26.57 | 85.9 | 21.5 |
| a | 8 | 2.22 | 4.58 | 2.06 | 5.09 | [410] | 30.96 | 56.0 | 56.0 |
| a | 9 | 2.27 | 4.58 | 2.02 | 5.11 | [810] | 37.87 | 109.6 | 109.5 |
| a | 10 | 2.29 | 4.58 | 2 | 5.12 | [100] | 45.00 | 27.2 | 27.1 |
| b | 1 | 2.43 | 2.29 | 1.06 | 3.34 | [110] | 0.00 | 76.8 | 38.4 |
| b | 2 | 2.29 | 2.29 | 1 | 3.24 | [430] | 8.13 | 135.8 | 135.8 |
| b | 3 | 2.22 | 2.29 | 0.97 | 3.19 | [320] | 11.31 | 195.9 | 97.9 |
| b | 4 | 2.29 | 2.29 | 1 | 3.24 | [15 8 0] | 16.93 | 230.9 | 230.8 |
| b | 5 | 2.05 | 2.29 | 0.90 | 3.07 | [210] | 18.43 | 60.7 | 60.7 |
| b | 6 | 2.29 | 2.29 | 1 | 3.24 | [12 5 0] | 22.38 | 176.6 | 176.5 |
| b | 7 | 2.17 | 2.29 | 0.95 | 3.16 | [310] | 26.57 | 85.9 | 85.9 |
| b | 8 | 2.22 | 2.29 | 0.97 | 3.19 | [410] | 30.96 | 112.0 | 112.0 |
| b | 9 | 2.27 | 2.29 | 0.99 | 3.23 | [810] | 37.87 | 219.0 | 219.0 |
| b | 10 | 2.29 | 2.29 | 1 | 3.24 | [100] | 45.00 | 27.2 | 27.1 |
| c | 1 | 4.86 | 2.29 | 0.47 | 5.37 | [110] | 0.00 | 38.4 | 38.4 |
| c | 2 | 4.58 | 2.29 | 0.5 | 5.12 | [430] | 8.13 | 135.9 | 135.8 |
| c | 3 | 4.45 | 2.29 | 0.51 | 5.00 | [320] | 11.31 | 196.0 | 97.9 |
| c | 4 | 4.58 | 2.29 | 0.5 | 5.12 | [15 8 0] | 16.93 | 231.0 | 230.8 |
| c | 5 | 4.10 | 2.29 | 0.56 | 4.69 | [210] | 18.43 | 60.8 | 60.7 |
| c | 6 | 4.58 | 2.29 | 0.5 | 5.12 | [12 5 0] | 22.38 | 176.6 | 176.5 |
| c | 7 | 4.35 | 2.29 | 0.53 | 4.91 | [310] | 26.57 | 85.9 | 85.9 |
| c | 8 | 4.44 | 2.29 | 0.52 | 5.00 | [410] | 30.96 | 112.0 | 112.0 |
| c | 9 | 4.55 | 2.29 | 0.5 | 5.09 | [810] | 37.87 | 219.1 | 218.9 |
| c | 10 | 4.58 | 2.29 | 0.5 | 5.12 | [100] | 45.00 | 27.2 | 27.1 |

# Competing interests

## Fundings

This work was funded by the National Natural Science Foundation of China (grant number: 11802112).

## Author contributions

W. Wan performed simulation, data gathering and validation. The original manuscript was written by W. Wan and jointly revised by W. Wan and C.X. Tang. The project was carried out and supervised by C.X. Tang.